\let\includefigures=\iftrue
%
\let\useblackboard=\iftrue
%
%
\newfam\black
\input harvmac
\input psfig
\noblackbox
\includefigures
\message{If you do not have epsf.tex (to include figures),}
\message{change the option at the top of the tex file.}
\input epsf
\def\figin{\epsfcheck\figin}\def\figins{\epsfcheck\figins}
\def\epsfcheck{\ifx\epsfbox\UnDeFiNeD
\message{(NO epsf.tex, FIGURES WILL BE IGNORED)}
\gdef\figin##1{\vskip2in}\gdef\figins##1{\hskip.5in}
\else\message{(FIGURES WILL BE INCLUDED)}%
\gdef\figin##1{##1}\gdef\figins##1{##1}\fi}
\def\DefWarn#1{}
\def\figinsert{\goodbreak\midinsert}
\def\ifig#1#2#3{\DefWarn#1\xdef#1{fig.~\the\figno}
\writedef{#1\leftbracket fig.\noexpand~\the\figno}%
\figinsert\figin{\centerline{#3}}\medskip\centerline{\vbox{
\baselineskip12pt\advance\hsize by -1truein
\noindent\footnotefont{\bf Fig.~\the\figno:} #2}}
\bigskip\endinsert\global\advance\figno by1}
\else
\def\ifig#1#2#3{\xdef#1{fig.~\the\figno}
\writedef{#1\leftbracket fig.\noexpand~\the\figno}%
\global\advance\figno by1}
\fi
%

\def\smallfig#1#2#3{\DefWarn#1\xdef#1{fig.~\the\figno}
\writedef{#1\leftbracket fig.\noexpand~\the\figno}%
\figinsert\figin{\centerline{#3}}\medskip\centerline{\vbox{
\baselineskip12pt\advance\hsize by -1truein
\noindent\footnotefont{\bf Fig.~\the\figno:} #2}}
\endinsert\global\advance\figno by1}

\useblackboard
\message{If you do not have msbm (blackboard bold) fonts,}
\message{change the option at the top of the tex file.}
\font\blackboard=msbm10 scaled \magstep1
\font\blackboards=msbm7
\font\blackboardss=msbm5
\textfont\black=\blackboard
\scriptfont\black=\blackboards
\scriptscriptfont\black=\blackboardss

\else

\fi
%
\def\yboxit#1#2{\vbox{\hrule height #1 \hbox{\vrule width #1
\vbox{#2}\vrule width #1 }\hrule height #1 }}
\def\fillbox#1{\hbox to #1{\vbox to #1{\vfil}\hfil}}
\def\ybox{{\lower 1.3pt \yboxit{0.4pt}{\fillbox{8pt}}\hskip-0.2pt}}
%
%

\def\comments#1{}

\def\tA{\tilde A}

\def\p{\partial}

\def\eps{\epsilon}
\def\tphi{\tilde \phi}
\def\half{{1\over 2}}

\def\CF{{\cal F}}


\def\CV{{\cal V}}


\def\ap{\alpha'}

\def\II{\relax{I\kern-.10em I}}

\def\IZ{\relax\ifmmode\mathchoice
{\hbox{\cmss Z\kern-.4em Z}}{\hbox{\cmss Z\kern-.4em Z}}
{\lower.9pt\hbox{\cmsss Z\kern-.4em Z}}
{\lower1.2pt\hbox{\cmsss Z\kern-.4em Z}}
\else{\cmss Z\kern-.4emZ}\fi}
\def\IB{\relax{\rm I\kern-.18em B}}
\def\IC{{\relax\hbox{$\inbar\kern-.3em{\rm C}$}}}
\def\ID{\relax{\rm I\kern-.18em D}}
\def\IE{\relax{\rm I\kern-.18em E}}
\def\IF{\relax{\rm I\kern-.18em F}}
\def\IG{\relax\hbox{$\inbar\kern-.3em{\rm G}$}}
\def\IGa{\relax\hbox{${\rm I}\kern-.18em\Gamma$}}
\def\IH{\relax{\rm I\kern-.18em H}}
\def\II{\relax{\rm I\kern-.18em I}}
\def\IK{\relax{\rm I\kern-.18em K}}
\def\IP{\relax{\rm I\kern-.18em P}}

%

\def\inbar{\,\vrule height1.5ex width.4pt depth0pt}

\def\p{\partial}

\font\cmss=cmss10 
\def\IR{\relax{\rm I\kern-.18em R}}

%


%

\def\gs{g_s}
\def\lp10{\ell_p^{10}}
\def\lp11{\ell_p^{11}}
\def\R11{R_{11}}

\def\frac#1#2{{#1 \over #2}}

\def\bns{B^{{\rm NS-NS}}}

\def\ie{{\it i.e.}}
\def\cf{{\it c.f.}}

\hyphenation{Di-men-sion-al}

\def\np{{\it Nucl. Phys.}}
\def\prl{{\it Phys. Rev. Lett.}}
\def\pr{{\it Phys. Rev.}}
\def\pl{{\it Phys. Lett.}}
\def\atamp{{\it Adv. Theor. Math. Phys.}}

\def\mpl{{\it Mod. Phys. Lett.}}

\def\jhep{{\it J. High Energy Phys.}}
\def\ijmp{{\it Int. J. Mod. Phys.}}
\def\ncim{{\it Il Nuovo Cim.}}
\def\annp{{\it Ann. Phys.}\ (NY)}

\lref\triangulate{S.B. Giddings, E. Martinec and E. Witten,
``Modular invariance in string field theory,'' \pl\ {\bf B176}\
(1986) 362.}

\lref\petr{P. Ho\v{r}ava, ``Type IIA D-branes,
K-theory and matrix theory,''
\atamp\ {\bf 2}\ (1999) 1373; hep-th/9812135.  See also
the Strings '99 talk
``M-theory, Mach's principle, and tachyon
condensation on branes'' at
http://strings99.aei-potsdam.mpg.de/.}
\lref\wktwo{E. Witten, ``Overview of K-theory
applied to strings''; hep-th/0007175.}

\lref\wopen{E. Witten, ``Noncommutative geometry and
string field theory,'' \np\ {\bf B268}\ (1986) 253.}
\lref\sbgven{S.B. Giddings, ``The Veneziano
amplitude from interacting string field theory,''
\np\ {\bf B278}\ (1986) 242.}
\lref\nbopenone{N. Berkovits, ``Super-Poincar\'e
invariant superstring field theory,''
\np\ {\bf B450}\ (1995) 90; erratum in {\it ibid.},
{\bf B459}\ (1996) 439; hep-th/9503099.}
\lref\nbopentwo{N. Berkovits, ``A new approach to
superstring field theory,''
{\it Fortsch. Phys.}\ {\bf 48}\ (2000) 31; hep-th/9912121.}
\lref\barton{B. Zwiebach, ``Closed string
field theory: quantum action and the B-V master
equation,'' \np\ {\bf B390}\
(1993) 33; hep-th/9206084.}
\lref\sftgauge{W. Siegel and B. Zwiebach, ``Gauge string fields,''
\np\ {\bf B263} (1986) 105.}
\lref\closedpoles{S.B. Giddings and E. Martinec,
``Conformal geometry and string field theory,''
\np\ {\bf B278} (1986) 91.}
\lref\ocone{B. Zwiebach, ``Quantum open string
theory with manifest closed string factorization,''
\pl\ {\bf B256}\ (1991) 22.}
\lref\ocinterp{B. Zwiebach, ``Interpolating string
field theories,'' \mpl\ {\bf A7}\ (1992) 1079.}
\lref\octwo{B. Zwiebach, ``Oriented open-closed string
theory revisited,'' \annp\ {\bf 267}\ (1998) 193;
hep-th/9705241.}
\lref\wbsftone{E. Witten, ``On background-independent
open string field theory,'' \pr\ {\bf D46}\ (1992) 5467;
hep-th/9208027.}
\lref\wbsfttwo{E. Witten,
``Some computations in background-independent
off-shell string theory,'' \pr\ {\bf D47}\ (1993) 3405;
hep-th/9210065.}
\lref\liwitten{K. Li and E. Witten, ``Role of short
distance behavior in off-shell open string field theory,''
\pr\ {\bf D48}\ (1993) 8535; hep-th/9303067.}
\lref\sbsftone{S. Shatashvili, ``Comment on the
background independent open string field theory,''
\pl\ {\bf B311}\ (1993) 83; hep-th/9303143.}
\lref\sbsfttwo{S. Shatashvili, ``On the problems
with background independence in string theory'';
hep-th/9311177.}
\lref\cornalba{L. Cornalba, ``Tachyon condensation in large magnetic
fields with background independent string field thery'';
hep-th/0010021.}
\lref\okuyama{K. Okuyama, ``Noncommutative tachyon from
background independent open string field theory'';
hep-th/0010028.}
\lref\andreev{O. Andreev, ``Some computations of partition
functions and tachyon potentials in background independent
off-shell string theory''; hep-th/0010218.}
\lref\nofactor{D.Z. Freedman, S.B. Giddings, J.A. Shapiro
and C.B. Thorn, ``The nonplanar one-loop amplitude in
Witten's open string field theory,'' \np\ {\bf B287}\
(1987) 61}
\lref\nofactortwo{J.A. Shapiro and C.B. Thorn,
``Closed string-open string transitions and
Witten's string field theory,'' \pl\ {\bf B194}\ (1987) 43.}
\lref\khv{J. Khouri and H. Verlinde, ``On open/closed
string duality''; hep-th/0001056.}

\lref\korkutone{K. Bardakci, ``Dual models and
spontaneous symmetry breaking,'' \np\ {\bf B68}\ (1974) 331.}
\lref\korkuttwo{K. Bardakci and M.B. Halpern, ``Explicit
spontaneous breakdown in a dual model,'' \pr\ {\bf D10}\
(1974) 4230.}
\lref\korkutthree{K. Bardakci and M.B> Halpern,
``Explicit spontaneous symmetry breakdown in a dual model II:
N point functions,'' \np\ {\bf B96}\ (1975) 285.}
\lref\korkutfour{K. Bardakci, ``Spontaneous symmetry
breakdown in the standrad dual string model,''
\np\ {\bf B133}\ (1978) 297.}
\lref\callan{C. Callan and J. Maldacena,``Brane Death
and Dynamics from the Born-Infeld Action,''
\np\ {\bf B513} (1998) 198-212; hep-th/9708147}
\lref\tcond{A. Sen, ``Tachyon condensation on the
brane anti-brane system,''
\jhep\ {\bf 9808}\ (1998) 012; hep-th/9805170.}
\lref\nonbpsact{A. Sen, ``Supersymmetric worldvolume
action for non-BPS D-branes,''
, \jhep\ {\bf 9910}\ (1999) 008; hep-th/9909062.}
\lref\universal{A. Sen, ``Universality of the tachyon
potential,'' \jhep\ {\bf 9912}\ (1999)
027; hep-th/9911116.}
\lref\sztcond{A. Sen and B. Zwiebach, ``Tachyon
condensation in string field theory,''
\jhep\ {\bf 0003}\ (2000) 002; hep-th/9912249.}
\lref\susytcond{N. Berkovits, A. Sen and B. Zwiebach,
``Tachyon condensation in superstring field theory,''
\np\ {\bf B587}\ (2000) 147; hep-th/0002211.}
\lref\tomandemil{T. Banks and E. Martinec, ``The renormalization
group and string field theory,'' \np\ {\bf B294}\ (1987) 733.}
\lref\senredef{A. Sen, ``Some issues in non-commutative
tachyon condensation,'' \jhep\ {\bf 011} (2000) 035;
hep-th/0009038.}
\lref\gshiggs{A.A, Gerasimov and S.L. Shatashvili,
``Stringy Higgs mechanism and the fate of
open strings''; hep-th/0011009.}
\lref\gibbonshy{G. Gibbons, K. Hori, and P. Yi, 
``String fluid from tachyon condensation'';
hep-th/0009061.}

\lref\spinors{A. Sen, ``SO(32) spinors of
type I and other solutions on brane-anti-brane
pair,'' \jhep\ {\bf 9809}\
(1998) 023; hep-th/9808141.}
\lref\descent{A. Sen, ``Descent relations among
bosonic D-branes,'' \ijmp\
{\bf A14}\ (1999) 4061; hep-th/9902105.}
\lref\vortex{A. Sen, ``Vortex pair
creation on brane-anti-brane pair via
marginal deformation,''
\jhep\ {\bf 0006}\ (2000) 010; hep-th/0003124.}
\lref\antallumps{R. de Mello Koch, A. Jevicki, M. Mihailescu and
R. Tatar, ``Lumps and p-branes in open string field theory,'' \pl\
{\bf B482}\ (2000) 249; hep-th/0003031.}
\lref\lumps{J.A. Harvey and P. Kraus,
``D-branes as unstable lumps in bosonic
open string field theory,''
\jhep\ {\bf 0004}\ (2000) 012; hep-th/0002117.}
\lref\circlelumps{N. Moeller, A. Sen
and B. Zwiebach, ``D-branes as tachyon lumps in
string field theory,'' \jhep\ {\bf 0008}\ (2000) 039;
hep-th/0005036.}
\lref\chitwo{P. Kraus, J.A. Harvey, F. Larsen and
E.J. Martinec, ``D-branes and strings
as noncommutative solitons,''
\jhep\ {\bf 0007}\ (2000) 042; hep-th/0005031.}
\lref\dmr{K. Dasgupta, S. Mukhi and G. Rajesh,
``Noncommutative tachyons,'' \jhep\ {\bf 0006}\
(2000) 022; hep-th/0005006.}
\lref\mzone{J. Minahan and B. Zwiebach, ``Field theory
models for tachyon and gauge field dynamics,'' \jhep\ {\bf 09}\
(2000) 029; hep-th/0008231.}
\lref\mztwo{J. Minahan and B. Zwiebach, ``Effective tachyon dynamics
in superstring theory''; hep-th/0009246.}

\lref\yi{P. Yi, ``Membranes from five-branes
and fundamental strings from Dp-branes,''
\np\ {\bf B550}\ (1999) 214; hep-th/9901169.}
\lref\bhy{O. Bergman, K. Hori and P. Yi,
``Confinement on the brane,''
\np\ {\bf B580}\ (2000) 289; hep-th/0002223.}
\lref\stringfluid{G. Gibbons, K. Hori and P. Yi,
``String fluid from unstable D-branes'';
hep-th/0009061.}

\lref\ksam{V.A. Kostelecky and S. Samuel,
``On a nonperturbative vacuum for the open
bosonic string,'' \np\ {\bf B336}\ (1990) 263.}
\lref\jdavid{J. David, ``U(1) gauge invariance from open string
field theory,'' \jhep\ {\bf 10}\ (2000) 017;
hep-th/0005085.}
\lref\watimass{W. Taylor, ``Mass generation
from tachyon condensation for vector fields
on D-branes;'' hep-th/0008033.}
\lref\chione{J.A. Harvey, D. Kutasov and
E.J. Martinec, ``On the relevance of tachyons;''
hep-th/0003101.}
\lref\marginal{A. Sen and B. Zwiebach,
``Large marginal deformations in string field theory,''
\jhep\ {\bf 0010}\ (2000) 009;hep-th/0007153.}
\lref\gmsii{R. Gopakumar, S. Minwalla and A. Strominger,
``Symmetry restoration and tachyon conensation in open
string theory''; hep-th/0007226.}
\lref\natinothing{N. Seiberg, ``A note on background
independence in noncommutative gauge theories,matrix
model, and tachyon condensation''; hep-th/0008013.}
\lref\watiprivate{W. Taylor, talk given at Stanford
University, and private discussions.}

\lref\shatash{A. Gerasimov and S. Shatashvili, ``On Exact Tachyon
Potential in Open String Field Theory''; hep-th/0009103.}
\lref\garousi{M.R. Garousi, ``Tachyon
couplinsg on non-BPS D-branes and
Dirac-Born-Infeld action,''
\np\ {\bf B584}\ (2000) 284; hep-th/0003122.}
\lref\bergetal{E.A. Bergsheoff, M. de Roo, T.C. de Wit, E. Eyras
and S. Panda, ``T-duality and actions for non-BPS
D-branes,'' \jhep\ {\bf 0005}\
(2000) 009; hep-th/0003221.}
\lref\kmmone{D. Kutasov, M. Mari\~no and G. Moore,
``Some exact results on tachyon condensation in string
field theory'', \jhep\ {\bf 0010}\ (2000) 045; hep-th/0009148.}
\lref\kmmtwo{D. Kutasov, M. Mari\~no and G. Moore,
``Remarks on tachyon condensation in superstring field theory'';
hep-th/0010108.}

\lref\swnoncomm{N. Seiberg and E. Witten,
``String theory and noncommutative geometry,''
\jhep\ {\bf 9909} (1999) 032; hep-th/9908142.}
\lref\gms{R. Gopakumar, S. Minwalla and A. Strominger,
``Noncommutative solitons,'' \jhep\ {\bf 0005}\ (2000) 020;
hep-th/0003160.}
\lref\nctach{E. Witten, ``Noncommuting tachyons and string
field theory''; hep-th/0006071.}
\lref\fluxone{A.P. Polychronakos, ``Flux tube solutions
in noncommutative gauge theories''; hep-th/0007078.}
\lref\fluxtwo{D.P. Jatkar, G. Mandal and S.R. Wadia,
``Nielsen-Olesen vortices in noncommutative Abelian
Higgs model''; hep-th/0007078.}

\lref\estring{C.P. Burgess, ``Open string instability
in background electric fields,'' \np\ {\bf B294} (1987) 427.}
\lref\sst{N. Seiberg, L. Susskind and N. Toumbas,
``Strings in background electric field, space/time
noncommutativity and a new noncritical string theory,''
\jhep\ {\bf 0006}\ (2000) 021; hep-th/0005040.}
\lref\gmms{R. Gopakumar, J. Maldacena, S. Minwalla
and A. Strominger, ``S-duality and noncommutative
gauge theory,'' \jhep\ {\bf 0006} (2000)
036; hep-th/0005048.}

\lref\ggsdbi{M.B. Green and M. Gutperle,
``Comments on three-branes,'' \pl\ {\bf B377}\
(1996) 28; hep-th/9602077.}
\lref\tsdbi{A.A. Tseytlin, ``Selfduality of Born-Infeld
action and Dirichlet three-brane of type IIB superstring
theory,'' \np\ {\bf B469} (1996) 51; hep-th/9602064.}

\lref\dkps{M. Douglas, D. Kabat, P. Pouliot and S. H. Shenker, ``D-branes
and short distances in string theory," \np\ {\bf B485} (1997) 85;
hep-th/9608024.}

\lref\senflux{A. Sen, ``Fundamental strings in open
string theory at the tachyonic vacuum''; hep-th/0010240.}
\lref\senemail{A. Sen, private correspondence.}

\lref\bfss{T. Banks, W. Fischler, S. Shenker and
L. Susskind, ``M theory as a matrix model: a
conjecture,''\pr\ {\bf D55}\
(1997) 5112; hep-th/9610043.}
\lref\malda{J. Maldacena, ``The large-N
limit of superconformal field theories and
supergravity,'' \atamp\
{\bf 2}\ (1998) 231; hep-th/9711200.}
\lref\jpwhat{J. Polchinski, ``What is string
theory?'' in
{\it Fluctuating Geometries in Statistical Mechanics
and Field Theory}, F. David and P. Ginsparg, eds.}
\lref\sspert{S. Shenker, ``The strength
of nonperturbative effects in string theory'';
in {\it Cargese 1990: Proceedings,
Random Surfaces and Quantum Gravity}\ O. Alvarez,
E. Marinari and P. Windey, eds. (Plenum, NY, 1990) 191.}
\lref\affleckludwig{I. Affleck and A.W.W. Ludwig,
``Universal noninteger 'ground state degeneracy' in
critical quantum systems,'' \prl\ {\bf 67}\ (1991) 161.}
\lref\leastd{S. Elitzur, E. Rabinovici and G. Sarkissian,
``On least action D-branes,'' \np\ {\bf B541}\ (1999) 246;
hep-th/9807161.}

\lref\teqd{J.A. Harvey, S. Kachru, G. Moore
and E. Silverstein, ``Tension is dimension,''
\jhep\ {\bf 0003}\ (2000) 001; hep-th/9909072.}
\lref\ksconf{J. Kogut and L. Susskind, ``Vacuum
polarization and the absence of free quarks in
four dimensions,'' \pr\ {\bf D9}\ (1974) 3501.}
\lref\fsw{P. Fendley, H. Saleur and N.P. Warner,
``Exact solution of a massless scalar field with
a relevant boundary interaction;'', \np\ {\bf B430} (1994) 577;
hep-th/9406125.}
\lref\polyakov{A. Polyakov, ``Quantum gravity in two-dimensions,''
\mpl\ {\bf A2}\ (1987) 293.}
\lref\kpz{V.G. Knizhnik, A. Polyakov and A.B. Zamolodchikov,
``Fractal structure of 2-D quantum gravity,''
\mpl\ {\bf A3} (1988) 819.}
\lref\tscosmo{C. Schmidhuber and A.A. Tseytlin,
``On string consmology and the RG flow in 2-D field theory,''
\np\ {\bf B426} (1994) 187; hep-th/9404180.}
\lref\kpp{V.A. Kostelecky, M.J. Perry and R. Potting,
``Off-shell structure of the string sigma model,''
\prl\ (2000); hep-th/9912243.}
\lref\macdecay{J. Dai and J. Polchinski,
``The decay of macroscopic fundamental strings,''
\pl\ {\bf B220}\ (1989) 387.}
\lref\dgtalk{D. Gross, lecture given at the
Aspen Center for Physics Aug.-Sept. 2000 ``String dualities''
workshop, and private discussions.}
\lref\wittenna{E. Witten, ``Nonabelian theta vacua''
[or something likethat], \ncim\ .}
\lref\collision{J. Polchinski, ``Collision of macroscopic
fundamental strings,'' \pl\ {\bf B209}\ (1988) 252.}
\lref\hagedorn{J.J. Atick and E. Witten,
``The Hagedorn transition and the number of degrees of
freedom of string theory,'' \np\ {\bf B310}\ (1988) 291.}
\lref\shensuss{For discussions of this speculation see the talk by
S. Shenker at \break
http://quark.theory.caltech.edu/people/rahmfeld/Shenker/fs1.html
and D. Bigatti and L. Susskind,  "TASI
Lectures On The Holographic Principle,"
hep-th/0002044.}
\lref\padic{D. Ghoshal and A. Sen, ``Tachyon condensation
and brane descent relations in p-adic string theory,''
\np\ {\bf B584}\ (2000) 300; hep-th/0003278.}

\Title{\vbox{\baselineskip12pt\hbox{hep-th/0012081}
\hbox{SU-ITP-00/33}
\hbox{SLAC-PUB-8735}}}
{\vbox{
\centerline{Closed strings from nothing}}}
\smallskip
\centerline{Matthew Kleban$^{1}$, Albion Lawrence$^{1,2}$
    and Stephen Shenker$^{1}$}
\bigskip
\bigskip
\centerline{$^{1}${Department of Physics,
Stanford University, Stanford, CA 94305}}
\medskip
\centerline{$^{2}${SLAC Theory Group, MS 81, PO Box 4349, Stanford,
CA 94309}}
\bigskip
\bigskip
\noindent

We study the physics of open strings
in bosonic and type II string theories
in the presence of unstable D-branes.
When the potential energy of the
open string tachyon is at its minimum,
Sen has argued that only closed
strings remain in the perturbative spectrum.
We explore the scenario of Yi and of Bergman,
Hori and Yi,
who argue that the open string degrees
of freedom are strongly coupled and
disappear through confinement.   We discuss
arguments using open string field theory and
worldsheet boundary RG flows, which seem to indicate
otherwise.  We then describe a solitonic
excitation of the open string tachyon and gauge
field with the charge and tension of a fundamental closed
string.  This requires a double scaling limit where the
tachyon is taken to its minimal value and the electric
field is taken to its maximum value.  The resulting
flux tube has an unconstrained spatial profile;
and for large fundamental string charge, it appears
to have light, weakly coupled open strings living
in the core. We argue that the flux tube acquires a size
or order $\alpha'$ through sigma model and
string coupling effects; and we argue that
confinement effects make the light degrees of freedom
heavy and strongly interacting.

\Date{December 2000}

\newsec{Introduction}

The known nonperturbative definitions of
string- and M-theory vacua
\refs{\bfss,\malda}\ are remarkable yet conceptually
deficient.  They provide unitary, holographic descriptions
of quantum gravity; but locality is not manifest even
at macroscopic scales. Any sort of background independence
is also completely obscure.  Experience tells us that
the world is local down to microscopic scales.
Certainly the systems described in
\refs{\bfss,\malda}\ describe local spacetime physics;
however, such physics is hard to extract
from the field theoretic "boundary " variables used.
It seems plausible that there should  be another set of manifestly
local "bulk" variables.  Holography could then arise
via the fixing of some large gauge symmetry
of the theory \shensuss .  Such a description would be a major
step towards understanding holography in more
general backgrounds.

A natural candidate for such a theory
is string field theory; it is local down
to the string scale, and it has a
large gauge symmetry.  However, closed string
field theory, the natural guess for a theory of
quantum gravity, does not seem to be
complete as it stands.  The state of the art \barton\ is
an ``effective'' theory, defined
perturbatively;
vertices must be added at
every order in $\gs$, and large
regions of the moduli space of string diagrams
are integrated over in the construction of these vertices.
\foot{\cf\ \jpwhat\
for discussion of this issue with further
references.}  There is no simple explanation
within closed string field theory of the
rapid divergence of closed string perturbation
theory \sspert; the effect must be contained in the new vertices.
The physical D-branes which this
divergence signals must be added to the theory
by hand.

Open string field theory is in better shape.
Bosonic open string field theory
(and the NS sector of open superstring field
theory) has a classical Lagrangian which
can be written in closed form, and
which generates all perturbative diagrams
\refs{\wopen,\triangulate,\nbopenone,\nbopentwo}.
Closed strings appear in intermediate
channels of open string diagrams \closedpoles, and
the open string theory covers the moduli space
of closed string diagrams with at least one
hole \triangulate.
Perturbation theory diverges no more rapidly
than field theory; again, this reflects
the existence of D-branes, which contribute
effects of order $e^{-1/g_{open}^2}$ to
scattering amplitudes.

It remains to find D-branes as open string field
theory excitations, and to find a description
of closed string theories in the absence of
D-branes.
The study of unstable D-branes, pioneered
and largely carried out by A. Sen and his collaborators,
provides an attractive answer for type II and
bosonic string theories.  Begin with unstable
D-branes or D-brane-anti-D-brane
pairs, filling all of space.
These will contain open-string tachyons;
the endpoint of tachyon condensation is
the perturbative closed-string vacuum
(up to closed-string tachyons)
\refs{\tcond,\nonbpsact,\universal,\sztcond,\susytcond}.\foot{Early
work studying condensation of the open string tachyon
can be found in \refs{\korkutone,\korkuttwo,\korkutthree,\korkutfour}.}
D-branes appear
as configurations of the open-string tachyon
field in the closed string vacuum; the
open string vacuum remains at the core of the D-brane
\refs{\spinors,\descent,\vortex}.  Such configurations
have also been found using the tachyon potential calculated in
open bosonic string field theory
\refs{\antallumps,\lumps,\circlelumps,\chitwo,\dmr}.

We would therefore like to describe type II string
theory  via open string field theory living
on $N$ unstable D9-branes of type IIA,
or on $N$ $D9-\bar{D9}$-brane pairs in
type IIB  \refs{\petr,\wktwo}
(possibly as $N\to\infty$ \wktwo).
This raises a number of questions, particularly:
what are the dynamics of
open strings in the closed string vacuum?
How do closed strings emerge as weakly coupled
excitations?

We will attack these questions
by studying electric flux tubes with closed string
charge as the tachyon condenses, as in
\refs{\yi,\bhy,\chitwo,\stringfluid,\senflux}.
In section 2 we argue, following previous work, that that the
open strings in the closed string vacuum
are strongly coupled, at least in variables natural to world sheet conformal
field theory.  We will begin
with a selective survey of the literature surrounding this
issue.  In particular our claim appears to
contradict the worldsheet analysis
of \refs{\chione,\kmmone,\gshiggs},
and we discuss a resolution
of this contradiction by pointing out
order of limits implicit in various approaches.
We will also address the rather different
picture emerging from numerical calculations
in the level truncation scheme \refs{\jdavid,\watimass}\
and discuss the possible relationship via
nonlinear, nonlocal, field redefinitions.
Our conclusion is that, at least in the variables natural
to world sheet conformal field theory, the open strings
should disappear through a confinement
of the $U(1)$ gauge field on the unstable
D-brane (or the diagonal $U(1)$ gauge field
on a $D-\overline{D}$ pair) under which the
string endpoints are charged; this picture was
developed initially by P. Yi and his collaborators
\refs{\yi,\bhy}.
In section 3 we study solutions of the
tachyon-Born-Infeld action proposed in
\refs{\nonbpsact,\shatash,\kmmone,\kmmtwo},
and find a solution with the charge and tension
of a macroscopic closed string.  In this solution
the the electric field must be scaled
to its critical value (as in refs. \refs{\sst,\gmms};
see also \refs{\gibbonshy}),
and the profile appears completely unconstrained.
We study the physics of this solution in Sec. 4,
first discussing effects by which the flux tube
may be localized, and next addressing
the open string dynamics in the core of this
flux tube.  It appears that if the closed string
charge is large, then the open strings in the core
are light and weakly coupled.  We argue that
if the flux tubes are indeed localized, this will
not be the case.  The picture
we arrive at is that closed strings are a
collective excitation
in a strongly-coupled open string background,
following the scenario developed in \refs{\yi,\bhy}.

\newsec{The fate of open strings, or the dynamics of nothing}

{\obeylines\smallskip
{\it ``I ask for nothing, Master.''}
{\it ``And you shall receive it -- in abundance.''}
\hfill -- The Rocky Horror Picture Show
\bigskip}


\subsec{Arguments for strongly coupled open strings}

Let us begin with an unstable Dp-brane in type II or bosonic
string theory.  The perturbative spectrum of open strings
ending on the brane will include a tachyon, a $U(1)$
gauge field, massless scalars describing the transverse
fluctuations of the Dp-brane, and a tower of
massive open string modes.  The unstable brane decays
by condensation of the tachyon; at the minimum of the
tachyon potential, the theory has only
closed strings as perturbative excitations
\refs{\tcond,\nonbpsact,\universal,\sztcond,\susytcond}.

How do the open strings disappear as perturbative
excitations?  Sen \nonbpsact\ argued that the
Lagrangian for open string modes vanishes
at the endpoint of tachyon condensation, \ie\
\eqn\vanishing{
    L = f(\phi) \tilde{L}(\left\{\Psi\right\})
}
where $\phi$ is the tachyon, $f(\phi)$ vanishes
at the minimum of the tachyon potential, and $\Psi$ represents the
open string modes.  In particular,
worldsheet RG flow arguments \chione, and
calculations \refs{\kmmone,\cornalba,\okuyama,\andreev}
using the background-independent
string field theory proposed in \wbsftone,
show that the action for the $U(1)$ gauge field
with slowly varying field strength $F$ is as proposed
in \nonbpsact:
\eqn\senact{
    S = \int d^{p+1} x \frac{V(\phi)}{\gs}
        \sqrt{ - \det \left(g_{\mu\nu} - F_{\mu\nu}
        \right) }\ ,
}
where $V(\phi)/\gs$ is the potential energy for the tachyon and
$V(\phi) = 0$ in the closed string vacuum.

A related phenomenon is discussed in \gmsii.
A nontrivial
maximal rank, constant NS-NS B-field in Euclidean
space leads to a gauged $U(\infty)$ symmetry.
The authors of \gmsii\ propose that the
endpoint of tachyon condensation in this background is the ``nothing
state'' for which
the $U(\infty)$ symmetry is restored (the background
is gauge-invariant).  This symmetry forbids a nonvanishing
gauge kinetic term.  We adopt their coinage for our title.

A vanishing kinetic term generally signifies
strong coupling; there is little cost in action
for wild fluctuations over short distances
in space.  Said another way, we can rescale the fields
to maintain a nonvanishing kinetic term, but then the
interaction terms will have infinitely large coefficients.
We expect the open strings to
disappear via some sort of confinement mechanism
as advocated in \refs{\yi,\bhy}.  In particular the
endpoints of the open strings are charged
under the $U(1)$ gauge field in \senact.  If
the $U(1)$ confines, the the endpoints will be joined
by an electric flux tube (see Fig. 1).  In the following section
we will argue that this flux tube is a piece of fundamental string,
so that the open-string-plus-flux-tube configuration is
in fact a closed string \senflux.

\bigskip
\centerline{\vbox{\hsize=5.0in
\centerline{\psfig{figure=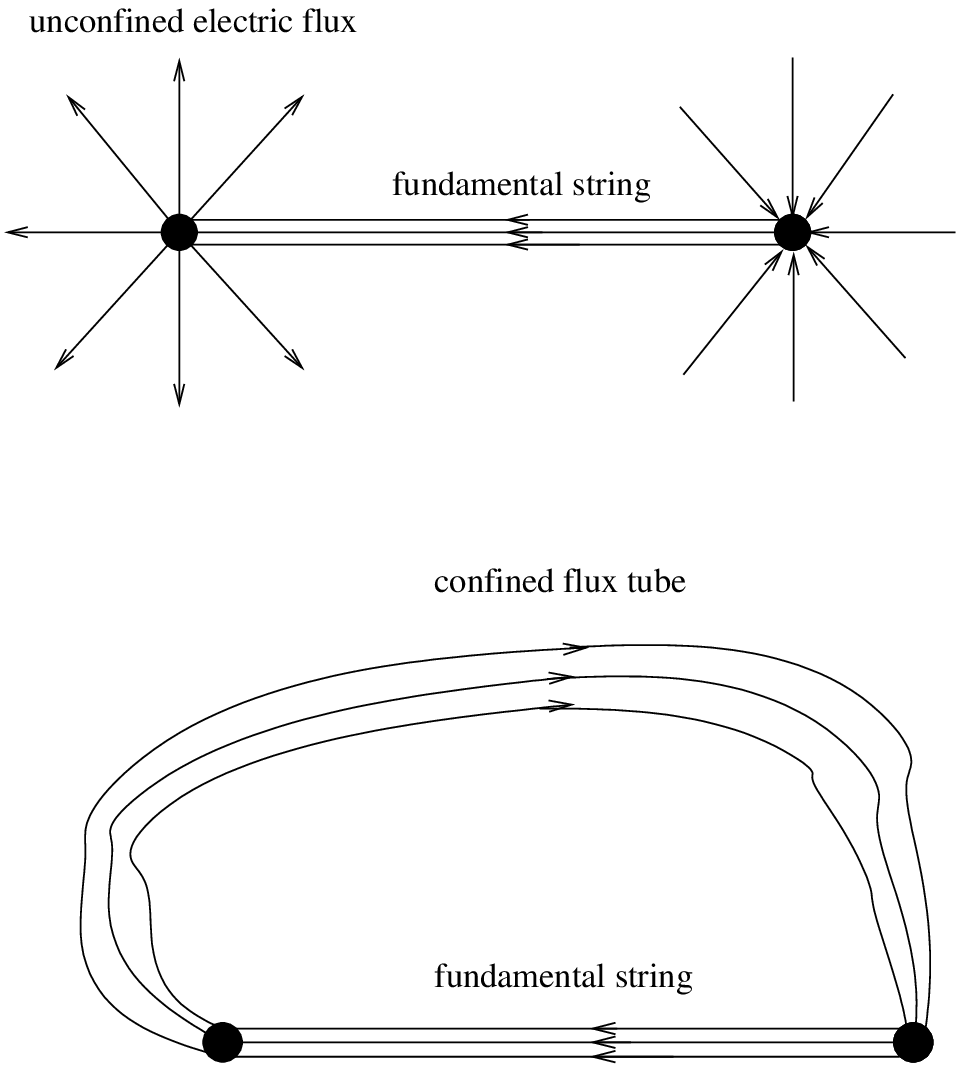}}
\vglue.7in
{\tenpoint Figure 1.
Confinement of an open string.}
}}
\bigskip

The tachyon is also strongly coupled
near its minimum.~\foot{We will persist
in calling $\phi$ ``the tachyon'' even
near the minimum of $V(\phi)$.}
Assume for the moment that $F=0$ and choose
coordinates on the field space of the tachyon for
which the tachyon kinetic term is the Lagrangian is
canonical:
$$ L_{kin} = \frac{1}{g_s} \left(\p\phi\right)^2\ . $$
The calculations in \refs{\wbsfttwo,\shatash,\kmmone,\kmmtwo}
show that near the closed string vacuum $\phi = 0$,
\eqn\canonpot{
    V(\phi) = - \frac{1}{\gs} \phi^2 \ln \phi^2\ + \ldots\ .
}
This precise form was also found using a scalar field
theory model of tachyon condensation in \refs{\mzone,\mztwo}.
(This is exact for the bosonic string but has
corrections for the superstring, as one can
see by performing the appropriate field redefinition
in \refs{\mztwo,\kmmtwo}.)  Since the second derivative
of this potential diverges logarithmically at $\phi = 0$,
it is tempting to claim that $\phi$ simply
decouples \shatash.  However, the higher derivatives
diverge as increasing inverse powers of $\phi$.
In any Feynman diagram describing tachyon scattering
near $\phi = 0$,
the diverging vertices overwhelm the vanishing propagators
for any ratio of propagators to vertices: a power law
divergence always overwhelms zeros from
arbitrary powers of inverse logs.

The mechanism proposed in \nonbpsact\ is
to set $V(\phi) = 0$ and treat the
gauge fields as Lagrange multipliers.
Upon integrating them out, the
charged states will be removed.  This
seems simpler than the strong coupling scenario, but
really is much the same.
In field theory, treating the gauge
fields as Lagrange multipliers is
the lowest order approximation in a strong-coupling
expansion.  Despite the simplicity of this
lowest-order solution, one
still understands the gauge fields as
fluctuating wildly.
Furthermore, $\phi$ will
fluctuate for any finite $g_s$; and
classical excitations of $\phi$ describe
D-branes and should be included in the theory.
In the end, we will have to face the full complications of
a strongly coupled system.

\subsec{Worldsheet analysis}

Equations \vanishing\ and \senact\
follow from world sheet and related string
field theory
analyses \refs{\chione,\shatash,\kmmone,\kmmtwo}.
These works study tachyon condensation by
adding a relevant boundary operator
\eqn\boundpert{
\eqalign{
    & S = S_0 + \delta S\cr
    &S_0 = \int_D d^2\sigma (\p X)^2\cr
    &\delta S = \oint_{\p D} dt T(X(t))
}}
to the tree-level open string worldsheet
(a disc with circumference $2\pi$)
and studying the infrared
fixed point of the induced RG flow.
The background-independent open string
field theory or ``boundary string field theory''
(BSFT) \refs{\wbsftone,\wbsfttwo,\liwitten,\sbsftone,\sbsfttwo}
constructs the effective action
using the worldsheet beta function
and the perturbed two-point functions.
The result is that the effective action
at the fixed points of these flows
is the boundary entropy as defined
in \affleckludwig.

Various choices of $T$ describe different endpoints.
The perturbation
\eqn\tacharray{
    T(X) = \lambda \sin (k \cdot X)
}
(for $k^2 < 1$) studied in \chione\ is integrable,
and the flow can be studied using Bethe-ans\"atz
techniques \fsw.  It describes an array
of lower-dimensional D-branes, or a D-brane on
a circle, in the IR limit $\lambda\to\infty$.
Similarly, a Gaussian perturbation
\eqn\tachdbrane{
    T(X) = u X^2
}
as studied in \refs{\shatash,\kmmone} describes
a single D-brane in the IR limit $u\to\infty$.
(More general quadratic interactions are
discussed in \liwitten.)
We note in passing that the tachyon profile
describing the D-branes is infinitely thin,
contradicting our expectations from the
low-energy effective action as well as explicit
calculations \refs{\antallumps,\lumps,\circlelumps}
using an approximation to
Witten's cubic string field theory, but in accord with our physical
expectations \dkps\ .
The calculations in \kmmone\ demonstrate that
the thinness of the D-brane arises
from higher-derivative corrections
which are not fully taken into account
in the low energy effective actions arising from in
either version of open string field theory.  This is an example of the
sensitivity
of results about string scale dynamics to truncation and field redefinition.

Finally, the constant perturbation
\eqn\tachconst{
    T(X) = \frac{a}{2\pi}
}
describes the closed string vacuum in the IR limit
$a \to \infty$.  Note that in these variables,
\eqn\redefone{
    \phi \propto e^{-a/2}
}
in the bosonic string for a
constant tachyon \refs{\shatash,\kmmone}.
A similar field redefinition in terms of
error functions can be performed for the
superstring \kmmtwo.

In the case of the constant perturbation
\tachconst, the terms in the tree-level effective
action up to two derivatives scale with
the tachyon as $V(\phi)$ near the
closed string vacuum.\foot{While
higher derivative terms scale to zero more
quickly -- \cf\ sec. 5 of \kmmone.}
As we discussed above, the vanishing of
the kinetic terms signals strong coupling.
The interaction terms also vanish.
But in the field theory limit of open string
diagrams, the vertices will vanish
but the propagators will diverge.
We expect the effective open string
coupling to diverge; roughly,
\eqn\roughcoupl{
    g_{eff}^2 = \frac{g_s}{V(\phi)}\ .
}

Two aspects of the discussion in \refs{\chione,\kmmone}
are apparently at odds with our
strong-coupling scenario.
First, one may study higher-loop amplitudes
in the presence of the perturbation \tachconst.
The authors of \kmmone\ argued that since
the perturbation is a constant, one might expect that each hole
is weighted with:
\eqn\wrongweighting{
    \tilde{g}_{eff}^2 \sim e^{-\oint a} =
        e^{-2\pi a}\ .
}
As $a\to\infty$ the holes will vanish, so the
effective open string coupling would be weak.

However we expect that on general grounds the correct
normalization of propagators
and vertices is given by the effective action
\vanishing,\senact\ extracted from
world sheet results.  One can already see at the
level of the four-point function that the integration
over moduli of the worldsheet will give a scaling different from
that implied by \wrongweighting.
For example, a representation of four-string
scattering is given by sewing together cubic vertices
and propagators using the cubic string field theory
in \wopen.  The relation would be a conformal
mapping along the lines of \sbgven: when two vertex
operators collide on the disc, the propagator in the
cubic SFT picture is getting long.  The constant tachyon
perturbation is not invariant under this conformal transformation;
and in the limit that the propagator is of length $L$,
the amplitude will not be suppressed by $e^{-a L}$.
A Riemann surface picture of the vertices and
propagators in BSFT has not yet been developed;
when it has, the scaling with $a$
of vertices and propagators should
be consistent with the effective action \senact.

A more serious objection to the strong coupling picture
is that the RG flows induced by the
perturbations \tacharray-\tachconst\
are nonsingular; there is no divergence or bad behaviour
at the level of the disc which would signal strong coupling.
The disc-level RG calculations seem to give
the right physics.  For perturbations
\tacharray\ and \tachdbrane\ describing D-branes
at the IR fixed points,
the open strings are bound by the perturbing potential
to the D-branes, and the D-brane tension is given exactly by the
tree-level boundary entropy.  For the constant
perturbation \tachconst, the effective action for
the open strings simply disappears.

Nonetheless, we still see
no contradiction with the strong-coupling
scenario.  To begin with, there is no reason for
the emergence of
strong coupling to appear as a divergence on the
disc.  For example, the worldsheet theory of perturbative
type II string theory in ten dimensions is
not singular as the dilaton is dialed to strong coupling;
rather, string diagrams get larger at each
order in perturbation theory.  Similarly, the
linear dilaton vacuum is an exact CFT even though it
contains a strong coupling region.

Secondly, if we take our effective action seriously,
at any value of $\phi$ away from zero, we
can scale $g_s$ to be much smaller that $V(\phi)$.
Thus, the disc amplitude accurately describes
physics arbitrarily close to the closed string vacuum
for arbitrarily small $g_s$.

Finally, when D-branes are present, we may self-consistently
allow the strings to end only on the D-branes.
The boundary entropy as measured by such worldsheets
must be the D-brane tension as proven in \refs{\leastd,\teqd}.
The mystery is why the strings seem to be bound
to the D-branes at tree level, when the strong coupling
picture suggests that the disappearance of
open strings far from the D-brane is described by
some sort of confinement mechanism.  For macroscopic
strings, a likely answer is that
the effective inverse coupling $V(\phi)/g_s$
can be treated as a dielectric constant
for the $U(1)$ gauge field, as in \ksconf.
The endpoints of strings are charged
and will energetically prefer to live in
a region with large dielectric constant.
Far from the D-branes there is no
gradient of the tachyon field; but the wavefunction
for the string endpoints will generally have support
intersecting the D-brane.  When the D-branes
are absent, we must resort to the strong coupling
picture to rid ourselves of open strings.

\subsec{Numerical calculations using cubic SFT}

The present excitement surrounding
open string field theory was in large measure triggered
by a calculation of the tension
of an unstable D-brane \sztcond\ using the open string
field theory of \wopen\ combined with the level truncation
scheme developed in \ksam.  Further work
on the low-energy effective action for
string fields in this scheme appears in
\refs{\marginal,\jdavid,\watimass}.  The results
of these calculations are again at odds
with the strong coupling picture, and
even with the effective action \senact.

Before addressing these apparent contradictions,
we should note that the RG calculations
of \refs{\chione,\shatash,\kmmone}\ are exact
and include all of the $\ap$ corrections; and
while the level truncation calculations
converge there is no understanding of
why they do.

That said,
the calculation of the tachyon potential energy
\refs{\sztcond,\susytcond}\
is classical and gives the correct answer
for the D-brane tension
to high accuracy.  If the
open strings are strongly coupled
in the closed string vacuum, why were we able to trust the
calculation close to
minimum of the tachyon potential?
This problem is identical to that of
the calculation of the D-brane tension in
BSFT.  Again, there exists
an order of limits such that at any point
in field space arbitrarily close to the
minimum of the tachyon potential, one may
dial the string coupling to be sufficiently small
so that the effective coupling $V(\phi)/\gs$
is weak.

An apparently genuine disagreement with
our scenario and with the form \senact\ of
the effective action arises
from calculations of the effective action
for the tachyon and $U(1)$ gauge field, using
the string field theory of \wbsftone.  At finite order
in the level truncation expansion, the
gauge field acquires a mass
\refs{\marginal,\jdavid,\watimass} indicating that gauge invariance
is broken and the effective action \senact\ is
incorrect.  Furthermore there is no
indication as yet that the string fields
are strongly coupled.  Results for
the coefficient of the gauge field kinetic term
have appeared in \jdavid; and while this
coefficient is decreasing at each order
in the level truncation approximation,
it is not decreasing very rapidly.

Nonetheless, we expect the cubic SFT and BSFT to
describe the same theory, using very different
coordinates on field space.  In the open string vacuum,
there are known finite field redefinitions
\refs{\kpp,\jdavid,\watimass,\kmmone}\ which
relate the fields used in cubic
SFT to fields with standard gauge transformations
as would appear in \senact.
In Appendix A we show that the field redefinition
between these two versions of open SFT are singular
in the closed string vacuum.
We also show that
if the effective action in BSFT variables
is \senact, then near the closed string vacuum
the field redefinition will lead to
large interaction terms in the action
in the cubic SFT variables.  We should
also note that the field redefinitions
could be highly nonlocal.  This is already
apparent near the open string vacuum as
shown in Appendix A of \kmmone.

Since the field redefinitions are singular
at the closed string vacuum, there
may be a preferred set of coordinates
there in which physics is weakly coupled.  Sen
has advocated that the open string degrees
of freedom are like angular variables at the
origin in polar coordinates \senredef.
At this point the reader may ask why
we should describe the closed string
vacuum via open string field theory;
the natural physical fields in the
closed string vacuum are, after all, closed strings.
As before, we emphasize that open SFT is the
best candidate for a fundamental description
of the theory.  It is certainly not excluded that a weakly coupled open string
description is valid in the physical vacuum,  with some as yet poorly
understood classical mechanism responsible for the removal of open string
states from the spectrum.

Given the rapid convergence of
numerical results, the results
in \refs{\marginal,\jdavid,\watimass}\ certainly
deserve further study.  But we see
no contradiction yet with the picture
developed in \refs{\nonbpsact,\shatash,\kmmone};
so we will continue to assume that this
latter picture of tachyon condensation
is correct, although it may not be the only picture.

\newsec{Electric flux tube solutions}

As promised, we will search for an
excitation of open string fields around
the closed string vacuum, with the mass
and charge of a fundamental string.   An obvious obstacle
confronts us--solitons typically have mass $\sim m_s/g_{open}^2$ while
we seek a solution with mass~$\sim m_s$.   We shall see, following
previous work, how to resolve this.

We will study an unstable non-BPS brane
in either type II or bosonic string theory.
We will focus on
the $U(1)$ gauge field $A_\mu$ and the neutral
scalar ``tachyon'' $\phi$.
There is no good reason to ignore the higher
open string oscillator modes;  indeed we will argue
that such modes become light.  These should
appear via higher-derivative corrections
to the Lagrangian we write down.

\subsec{The action}

For slowly varying $U(1)$ fields, we assume following
\refs{\swnoncomm,\nonbpsact} that the
kinetic term is constructed with the open-string
metric.  We use the following form:
\eqn\kinetictachyon{
    S_{\phi, kin} =
    \frac{1}{\gs}\int d^{p+1} x \sqrt{-\det \left(g_{\mu\nu} + \CF_{\mu\nu}
    \right)} \left( - G_S^{\mu\nu}
    \p_\mu\phi\p_\nu\phi\ - V(\phi) \right) .
}
Here $\CF = 2\pi\ap F - \bns$, and $G_{S,\mu\nu}$ is the symmetric part of
$(g_{\mu\nu} + \CF_{\mu\nu})^{-1}$.  We use the tachyon
lagrangian computed in \refs{\wbsfttwo,\shatash,\kmmone,\kmmtwo},
choosing field variables for which the tachyon
kinetic term is canonical; $V(\phi)$ has the form
\canonpot\ near $\phi = 0$.

\subsec{General cylindrically symmetric solution}

We will work in cylindrical coordinates
$(t,z,\rho,\Omega)$ in $p+1$ dimensions,
where $z$ is the axial direction, $\rho$ the radial direction,
and $\Omega$ the the $(p-2)$ remaining angular coordinates.
We will look for a purely electric, axially symmetric solution:
$\vec{E} = E(\rho) \hat{z}$, $F_{ij} = 0$ for
$0 < (i,j) \leq p$;
and $\phi = \phi(\rho)$.

Using our cylindrically symmetric ans\"atz
in the action \kinetictachyon,
we find that:
\eqn\bicylact{
    S = - \frac{1}{\gs} \int d^{p+1} x
    \sqrt{1-(2\pi\ap E)^2} \left( V(\phi) + (\partial_\rho \phi)^2
        \right)
}
The canonical momentum for the electric field is:
\eqn\bimom{
    D_z = \frac{(2\pi\ap)^2 E}{\gs \sqrt{1-(2\pi\ap E)^2}} \left( V(\phi)
        + (\partial_\rho \phi)^2 \right)
}
and the Hamiltonian as a function of $E$ is:
\eqn\biham{
    H = \int d^{p} x \frac{1}{\gs \sqrt{1-(2\pi\ap E)^2}}
        \left( V(\phi)
        + (\partial_\rho \phi)^2 \right) =
    \int d^{p}x \frac{D}{(2\pi\ap)^2 E}\ .
}
We impose the charge quantization condition:
\eqn\bicq{
    Q = \int \rho^{p-2}d\rho D = k\ .
}
If we wish the tension to be $\frac{k}{2\pi\ap}$ as
it must be for $k$ fundamental strings, then
eq. \biham\ means that we must scale $2\pi \ap E = 1$,
which is its maximal value \estring.  This limit allows us to have a very
light soliton.

We can find a solution by varying the Lagrangian plus
the charge constraint:
\eqn\varywcon{
        \delta \left[ S + \lambda \left(Q - k\right)\right]\ ,
}
where $\lambda$ is a Lagrange multiplier.
Varying with respect to $E$, we find that $E$ is a function
of $\lambda$ and is therefore a constant.
$\gamma = \sqrt{1 - 2\pi \ap E}$ scales out
of eq. \bicylact, so that the equation
of motion for $\phi$ is independent of $E$.

\subsec{Two solutions}

\vskip .3cm
\noindent{\it Flux tube with tachyon profile}
\vskip .2cm

Following \refs{\antallumps,\lumps,\circlelumps,\chitwo,\dmr},
we assume that eq. \bicylact\ has a cylindrically
symmetric lump solution.  Even without a specific solution,
we can study the energetics by scaling arguments.
The tachyon potential
has been computed at tree level, so the only dimensionful
parameter in the solution will be $\ap$ which
we set to one.
The transverse integral
\eqn\transvnorm{
        \beta = \int \rho^{p-2}d\rho \left(V(\phi) +
                (\p_\rho\phi)^2\right)
}
will be order one in these units (possibly times
some integer $n$ if there are several lumps).  We can solve
eq. \bicq\ for $E$ in terms of $\beta,\gs$.
The energy per unit length (along $z$) of the solution is:
\eqn\lumpenergy{
        H = \frac{1}{\gs} \sqrt{ \beta^2 + \gs^2 k^2}
}
which is the tension of the bound state of a D-string
and $k$ fundamental strings.  This should come as no
surprise since it is believed
\refs{\antallumps,\lumps,\circlelumps,\chitwo,\dmr} that
lump solutions of the tachyon should correspond to
D-branes.

If we expand the
Born-Infeld action to quadratic order in $F$,
we recover an action almost identical to that
discussed in \ksconf\ as a model of confinement.\foot{
The difference is that in that work, the
function of $\phi$ multiplying $F^2$ differed
from $V(\phi)$ in such a way that
electric flux tubes emanating from
a point source were energetically
preferred to a spherically symmetric configuration
of the electric field emanating from the same source.
This is not guaranteed by the Maxwell limit
of the Lagrangian at hand.}
$V(\phi)$ appears as a field-dependent
dielectric constant.  Although $E$ is constant
everywhere, the canonical field momentum
is proportional to $V$ and dies off away
from the flux tube; thus the energy and
fundamental string charge are finite.

\vskip .3cm
\noindent{\it Fundamental string solution}
\vskip .2cm

It is clear from eq. \biham\ that if
we wish a solution with fundamental string
tension, we must work with the full Born-Infeld
action in the limit $\gamma\to 0$.
In this limit, we must also scale the tachyon
to keep $H$ and $\vec{D}$ finite.
We will take the double scaling limit:
\eqn\dscaling{
\eqalign{
    \sqrt{1 - (2\pi\ap E)^2} \sim
    -\frac{\epsilon^2\ln\epsilon^2}{\gs k}\cr
    \phi \sim \epsilon g(\rho)\ .
}}
As $\epsilon\to 0$, the tachyon equations
of motion vanish identically.  $g$ is
constrained only by charge quantization:
\eqn\bicqtwo{
    \int \rho^{p-2}d\rho g^2 + (\p_\rho g)^2/\ln \epsilon^2
    = 1\ .
}
Thus there is a continuous infinity of flux tube configurations,
as we will discuss below.
These results are implicit in refs. \refs{\bhy,\chitwo}.

\newsec{Physics of the Scaling Limit}

In order to study the properties of open string fields,
we will study the action for a generic open string
scalar field $\chi$.
Following the arguments in
\refs{\swnoncomm,\nonbpsact,\kmmone} we use
the Lagrangian
\eqn\genericscalar{
    S = \frac{1}{g_s} \int d^{p+1}x V(\phi)
    \sqrt{-\det \left(g_{\mu\nu} + F_{\mu\nu}\right)}
    \left(G_S^{\mu\nu}\p_\mu\chi \p_\nu\chi - m^2 \chi^2
    + \lambda \chi^3 + \ldots\right)\ .
}
Here $m^2 \sim M_s^2$ and $\lambda \sim 1$.  If we scale
$g_s$ out of the kinetic term, the cubic coupling will
be proportional to the open string coupling
$g_{open} =\sqrt{g_s}$.

In order to study the physics of open
string fluctuations around our flux tube
configuration, we expand \genericscalar\
in small fluctuations,
and rescale the fields and coordinates so that the kinetic
terms are all of order 1 as $\epsilon\to 0$ in
eq. \dscaling.  The effective mass and coupling
of the string theory will appear as the coefficient
of quadratic and higher-order terms in the effective
action for the string modes.  In particular, if
interaction terms become large then
we can conclude that the system is
strongly coupled.

In the open string vacuum, we know that the limit
$2\pi\ap E\to 1$ will send the open string
coupling to zero {\estring\sst\gmms}.
On the other hand, if we scale the tachyon
we expect the effective coupling
to be $g_{eff}^2 = \gs/V(\phi)$.
These two effects compete in the double scaling limit.

\subsec{The open string vacuum}

To check our formalism and as a warm up, we will first expand
the action around the open string vacuum in the
limite $2\pi \alpha' E \to 1$.  We define $\phi$ so that
$V(\phi) = 1 - \phi^2 + \ldots$ and we scale $\phi$
to zero independently of $1-(2\pi\alpha'E)^2$.

As before we scale
$$ 1 - (2\pi\alpha' E)^2 \sim \epsilon^4\ . $$
In this limit, the Lagrangian for $\chi$ is:
\eqn\openvaclag{
    S = \frac{1}{g_s} \int dt dz d^{p-1}x
    \left( \frac{1}{\eps^2} \left[ (\p_t\chi)^2
        - (\p_z\chi)^2 \right]
    - \eps^2 \left[ (\vec{\nabla}\chi)^2 + m^2 \chi^2
        + \lambda \chi^3 \right]\right)
}
where we have suppressed finite constants (factors of $2\pi$ and so on).

We wish to rescale $\chi$ and the coordinates $t,z,\vec{x}$,
so that the kinetic terms are independent of
$\epsilon$.
We write the most general rescalings:
\eqn\scalings{
\eqalign{
    & \chi = \eps^{\delta}\tilde{\chi}\cr
    & (t, z) = \eps^{\beta}
        (\tilde{t},\tilde{z})\cr
    & (x_1,\ldots,x_{p-1}) = \eps^{\beta + 2}
    (\tilde{x}_1,\ldots,\tilde{x}^{p-1})\cr
}}
The scaling of $\vec{x}$ differs from that of $t$ and $z$
due to the anisotropy of the open string metric.

A canonical ($\eps$-independent) kinetic term
requires:
\eqn\condition{
    \delta = 1 - {p - 1 \over 2} \left(
    \beta + 2 \right)
}
$\beta$ is not yet fixed; it rescales
the time coordinate and thus the energy scales.
We set $\beta = 0$, thus measuring energy in
units of $M_s$.

We have simply reproduced the NCOS scaling.
The scalar mass term scales as:
\eqn\chimass{
    m^2 \epsilon^4 \tilde{\chi}^2\ ;
}
if $m \sim M_s$, then the mass in \chimass\
is of order $M_{s,eff} \sim \epsilon^2$
as in NCOS theory.
The cubic term in $\chi$ scales as:
\eqn\chiscalingE{
    \tilde{\lambda} \tilde{\chi}^3
    \sim \epsilon^{6 - p}\tilde{\chi}^3 .
}
The coupling has mass dimension $(5-p)/2$.
To find the dimensionless coupling, we should
multiply this by
$M_{s,eff}^{(p-5)/2}= \eps^{p-5}$, as $M_{s,eff}$
is the natural mass scale in the theory.
The dimensionless coupling
$$\tilde{\lambda} M_{s,eff}^{(p-5)/2} \sim \eps$$
which is the expected scaling for the open string coupling \refs{\estring,
\sst,\gmms}.  Finally, if we write the line element for the open string metric
$$ ds^2 = G_{\mu\nu}^{open}dx^\mu dx^\nu $$
and rescale the coordinates so that:
$$ ds^2 = (\alpha'_{eff})
    \eta_{\mu\nu}d\tilde{x}^\mu d\tilde{x}^\nu\ ,
$$
the the rescaling is precisely that in \scalings.

\subsec{The fundamental string solution}

For the solution with $k$ fundamental strings in the
closed string vacuum,
the scaling \dscaling\ of the tachyon potential
keeps the dimensionless coupling of the $\chi$ finite.
In this scaling limit, the action \genericscalar\ is:
\eqn\genericscaling{
    S = \int d^{p+1}x g(\rho) \left(
    k \left[ (\p_t\chi)^2 - (\p_z \chi)^2\right]
    -k\left(\frac{-\eps^2\ln\eps^2}{g_s k}\right)^2
    \left[(\vec{\nabla}\chi)^2 + m^2\chi^2
    + \lambda \chi^3 + \ldots \right]\right)\ ,
}
where we have ignored the subleading
$\eps^2g^2\ln g^2$ term in $V(\eps g)$.
First, in order to make the kinetic term isotropic, we rescale
$\vec{x}$:
\eqn\newcoords{
    \vec{x} = \frac{-\eps^2\ln\eps^2}{g_s k}
    \vec{\tilde{x}}\ .
}
Next, in order to make the kinetic terms canonical, we
rescale $\chi$:
\eqn\newfield{
    \chi = \left(\frac{-\eps\ln\eps}{g_s k}\right)^{(1-p)/2}
        \frac{1}{\sqrt{g(\rho)k}}\tilde{\chi}\ .
}
where we have assumed $g(\rho)$ varies slowly.

In this limit the mass still scales as $\sqrt{1-(2\pi\alpha'E)^2}$:
\eqn\masspart{
    S_{mass} \sim \int d^{p+1}\tilde{x} \frac{-\eps^2\ln\eps^2}{g_s k}
        m^2 \tilde{\chi}^2
    = \int d^{p+1}\tilde{x} (M'_{eff})^2 \tilde{\chi}^2\ .
}
The cubic term scales as:
\eqn\interactpart{
    S_{cubic} \sim \int d^{p+1}\tilde{x}
    \frac{1}{\sqrt{g(\rho)k}} \left(
    \frac{-\eps^2\ln\eps^2}{g_s k}\right)^{(5-p)/2}
    \tilde{\chi}^3 =
    \int d^{p+1}\tilde{x} \tilde{\lambda}\tilde{\chi}^3\ .
}
The dimensionless coupling scales as:
\eqn\dimlesscoupling{
    G_{open}\sim \tilde{\lambda}(M_{eff}')^{(p-5)/2}
    = \frac{1}{\sqrt{g(\rho)k}}
}
Recall that $\int d^{p-1}\vec{x} g = 1$.
Thus if the flux tube has transverse
size $l_s$, $g\sim 1$ inside the flux tube
and the open string modes are weakly coupled
at large $k$.  Note that $G_{open}$
does not depend on $g_s$. Outside the flux tube,
$g$ vanishes
and the background is truly strongly coupled.

We are now left with two puzzles.  First,
can we remove the continuous infinity of
possible profiles?  Secondly, for $k \gg 1$
fundamental strings, how do we explain
the apparent appearance of light,
weakly coupled open string degrees of
freedom?

\subsec{Localization of flux tubes}

The apparent degeneracy of flux tube profiles is
serious, as such degrees of freedom do not
exist for the fundamental string we are attempting
to describe.\foot{Additional
discussion of this issue can be found in
refs. \refs{\chitwo,\senflux \bhy}.}
However, the solution above
is naturally subject to both sigma model
corrections and quantum corrections.
Some combination of
these two could  localize the flux tube.

One might think that the natural scale
at which sigma model corrections (namely,
corrections from higher-derivative terms
in $\phi$ and $F$) would become important
is $\ell_{s,eff}$.  If we assume
cylindrical symmetry, however, this
is not the case: the characteristic
scale of transverse spatial fluctuations
is still $\ap$.  One can see this
by inserting a solution $\phi = \epsilon g(\rho)$
into some higher-derivative term and comparing
it to the kinetic energy.   Only terms with $\rho$
derivatives will contribute.  The open string metric
in the $\rho$ direction does not scale with
$\sqrt{1 - (2\pi\ap E)^2}$, so the derivative
expansion for these terms is an expansion
in $\ap$.

The other option, explored in $\bhy$,
is that nonperturbative corrections
arising from $D(p-2)$-brane condensation
localize the flux tube.  The claim is that
this condensation confines the open string
degrees of freedom, and any flux tube has
a width naturally associated with the
confinement length.  In \bhy\
such an effect was argued for $D2-\bar{D2}$
pairs as arising from D0-brane instantons
stretching between the two $D2$-branes.
In this case the flux tube appears
infinitely thin.  Given the above discussion
of $\ap$ corrections, it is plausible that
such corrections smear out the solution
out to width $\ap$.

The authors of \bhy\ similarly argue for such an effect
in $D3-\bar{D3}$ systems from type IIB S-duality.
Here D-strings stretching between the D3-branes
are monopoles magnetically charged under the
diagonal $U(1)$.  One may use S-duality to argue that
these condense and confine electric flux, but
the string coupling is strong in the
S-dual picture so this description is heuristic at best.

In the end our belief is that some
combination of strong quantum effects
and $\ap$ corrections should lead to
a flux tube profile with width of order $\ap$.
Clearly our understanding is unsatisfactory
and the subject deserves further study.

\subsec{Weakly coupled open strings}

The results for the open string spectrum and coupling
in the core of the flux tube
are very confusing.
If the naive estimates are correct and the
flux tube indeed corresponds to $k$ fundamental strings,
then for $k\gg 1$ there should be a Hagedorn density
of light, weakly coupled excitations
{\it in addition} to the spatial fluctuations
of the fundamental strings (which are
captured by the collective coordinates
of the flux tube).  In particular these additional
weakly coupled states
would affect the results of perturbative closed string
scattering, even at low energies.

If the flux tubes are spread
out over a transverse volume
of order $k\ell_s^{p-1}$, then
the effective coupling $1/\sqrt{k g(\rho)}$
is order one or higher and it is plausible
the open strings disappear from the spectrum
in the core of the tube.  If the
flux tube is small, of order $\ell_s$ in size,
then the Compton wavelength of the light
open string modes is much larger than the
width of the flux tube by a factor of $1/\eps^2$.
These light modes prefer to stay within the flux tube,
but this costs an uncertainty principle energy of order
$$M_s \sim M_{s,eff}/\epsilon^2 \gg M_{s,eff}\ .$$
So these light excitations are pushed to energies
much higher than their characteristic string scale.
Since the coupling is of order $1/\sqrt{k}$,
$M_s$ is nonperturbatively larger than $M_{s,eff}$
if $\epsilon \ll 1$, and
there is no reason to assume the modes at energies
$M_s$ are weakly coupled.

However, quantum fluctuations generally keep us from setting $\phi$
strictly to its minimum; this would be reflected
in a minimal value for the scaling parameter $\epsilon$.
Nonetheless, we can see that $\epsilon$ will go to
zero in the limit of vanishing $g_s$.  Upon examining
the Schr\"odinger equation for the zero mode $\phi_0$ of $\phi$,
it is clear that the wavefunction of the ground state
should scale as $g_s^{1/2} (\ln g_s)^{\delta}$.
If we assume this is a minimum value for $\phi$
and plug this into $V(\phi)$, we find that
the minimum value for $\epsilon$ is
$$g_s = \epsilon^2 f(\ln \epsilon)$$
And so, up to logarithms,
$$ \ell_{s} = g_s \ell_{s, eff}\ . $$
Thus, for sufficiently small $g_s$
the energies of the
of the localized light modes in the flux tube
are nonperturbatively larger
than their masses because the effective open string coupling is
of order $1/\sqrt{k}$. At
such energies there is no longer any reason
to assume that the open strings interact weakly.

At any rate, while we have unfortunately placed
the open strings out of any
calculational control, we no longer
see any paradoxes arising from our flux
tube solution and so from our strong-coupling
picture of open string dynamics in the closed string
vacuum.

\newsec{Conclusions}

To take stock, let us consider the pros and cons of
studying closed strings as solitons in tachyon
condensed open string theory (as
opposed to keeping explicit closed string degrees of freedom).

First, the pros:  in contrast to
closed-string field theory,
open-string field theory
may be nonperturbatively well defined,
as well as bulk gauge covariant.  Within
this framework, the strongly coupled,
confining picture introduced in \refs{\yi,\bhy}
provides a natural explanation for the vanishing of
typical open string states
after tachyon condensation \bhy.   Begin with a rather
large open string when the
tachyon vev is in the open string vacuum.
The open string state consists of two
opposite charges joined by a section of fundamental string,
with unconfined flux flowing off the charged ends.   As
the tachyon rolls, the $U(1)$ becomes confining, and the flux
squeezes into a flux tube connecting the two charged ends.
But in our scaling limit, the fundamental string has the same
tension and charge as the flux tube, and so one really has
a closed loop of flux,  i.e., a closed string
state\foot{Ashoke Sen has discussed
similar picture \refs{\senflux,\senemail}}.

Now the cons. Basically nothing is calculable; the
degrees of freedom are strongly coupled, and
stringy rather than field theoretic (in contrast to
the background-dependent descriptions of \refs{\bfss,\malda}).
Such a messy description almost cries out for new,
more manageable variables.  Sen has made an intriguing analogy to
polar coordinates \senredef.  If we take a field theory with two
real scalar fields, we can rewrite the theory in polar
coordinates on field space:
\eqn\polar{
    S = \frac{1}{\hbar}\int d^d x
    \left[ (\p r)^2 + r^2 (\p \theta)^2\right] \ .
}
$\theta$ gets strongly coupled near the origin, but the
Cartesian coordinates are still weakly coupled
at small $\hbar$.  In the present case, the
perturbative excitations at $V(\phi) = 0$
are closed strings.  Unfortunately their nonperturbative
dynamics is problematic.  Still,
one might expect by this analogy that any description using
explicit open string fields --
for example either the cubic or the background independent
SFT -- would be analogous to the angular variables in polar
coordinates, since the closed string states must disappear.
Certainly the kinetic terms appear to vanish
in the closed string vacuum for string fields fields in BSFT.
We should note that it is still possible that the
cubic SFT variables are well behaved at $V(\phi) = 0$.
Although the coefficients of the gauge kinietic
term are decreasing at the lowest few orders
of the level truncation expansion \jdavid,
they are not converging as quickly as the results for
the D-brane tension \sztcond, and there
is no indication as yet that they vanish 
\refs{\jdavid,\watiprivate}.
If the kinetic terms and the mass of the gauge field
are of order one in units of $M_s$, and if the
interaction terms are finite, then the variables
natural to the cubic SFT will be perfectly good.
One still requires an explanation of the disappearance
of perturbative open string fields despite the finite mass
and kinetic terms.  One possibility is that the
poles disappear due to higher-derivative terms,
as occurs at low orders in the level truncation scheme
\ksam\ and in the field theory of the
p-adic string \padic.

To summarize: we have examined the construction
of macroscopic
closed strings from open string degrees of
freedom, an excitations of the closed
string vacuum.  It appears that such
a construction is consistent with
the known physics of closed strings,
but we still have no computational control
over such solutions.

There are a number of issues that deserve to addressed
at this point.  First,
our flux tube solutions are highly quantum-mechanical.
A typical classical field configuration has
field strengths of order $1/\hbar$.
The field strengths in the core of the flux
tubes we study are of order
$\hbar^0$.\foot{We would like to
thank L. Susskind for stressing this point.}

Secondly, we do not understand how
$g_s$ emerges as the closed string coupling
in our scenario.
In our solutions, $g_s$ scales out
of the effective coupling of the open string
degrees of freedom.  It is not clear how
the coupling between flux tubes can
be $g_s$.  With more
control over this system,
we would be able to measure this coupling
by studying the probability for two
flux tubes to intersect and rejoin
as in Fig. 2.

\bigskip
\centerline{\vbox{\hsize=4.0in
\centerline{\psfig{figure=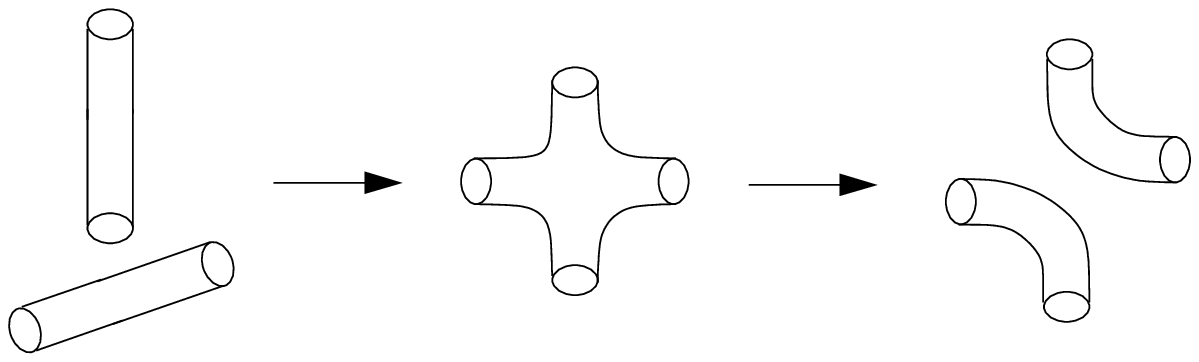}}
\vglue.4in
{\tenpoint Figure 2.
Fluxtube interaction.}
}}
\bigskip

Finally, a more ``mundane'' project suggested by sec. 2 is to better
understand the relationship between different
presentations of open string field theory.  The cubic SFT is tied
in a relatively intuitive way to worldsheet
geometry -- the moduli of a collection of
vertices and propagators with a given topology
cover the moduli of Riemann surfaces with that topology
\triangulate.  The quantum version of BSFT
is not well understood.  In view of its computational
power and conceptual elegance
in understanding classical tachyon condensation, this
picture deserves to be developed further.

\bigskip
\centerline{\bf{Acknowledgements}}
We would like to thank O. Bergman,
M. Douglas, D. Gross, J. Harvey, K. Hori, P. Kraus,
D. Kutasov, E. Martinec, J. McGreevy,
G. Moore, A. Sen, E. Silverstein, L. Susskind,
W. Taylor, N. Toumbas, N. Warner and P. Yi
for useful conversations and comments.
Some of this work was carried out at
the Aspen Center for Physics
``String Dualities'' workshop.
A.L. would also like to thank the
Rutgers High Energy Physics Group for
their hospitality while this work was initiated.
M.K. was supported
in part by Stanford University.
A.L. was supported in part by the DOE under
contract DE-AC03-76SF00515, and in part
by a DOE OJI grant awarded to E. Silverstein.
S.S. was supported in part by  NSF  grant
PHY-9870115.

\appendix{A}{Field redefinitions}

We would like to understand what field
redefinitions could relate the
variables natural to the cubic and
``background-independent'' string
field theories of \refs{\wopen,\wbsftone}.

As a guide, we
analyze the gauge invariance of the effective
action.  The gauge transformations
of the string fields depends on the choice
of coordinates on field space, which will depend
on the worldsheet regulator used to define the
propagator and interactions (\cf\ \kpp).  We can study
the gauge symmetries of the effective action
to get some insight into the field redefinitions
one needs to relate the calculations in
\refs{\sztcond,\susytcond,\jdavid,\watimass} to
the calculations in \refs{\wbsfttwo,\shatash,\kmmone}.
In general these field redefinitions will
be complicated and nonlocal and involve all of
the string fields.  We will study
local transformations of the tachyon
and gauge field only, which should suffice
for nearly static configurations after integrating
out the auxilliary fields.

Let $\tphi,\tA$ be the tachyon and gauge field
in the cubic open string field theory,
where $\tphi = 0$ is the open string vacuum.
At $\tphi = 0$, the
$U(1)$ gauge transformation is \refs{\sftgauge,\ksam}:
\eqn\sftgauge{
\eqalign{
    \delta\tphi = - \tA^\mu \p_\mu \Lambda + \ldots\cr
    \delta tA_\mu = \p_\mu \Lambda +\ldots\ ,
}}
where $\Lambda$ is the gauge parameter.
It is already clear that the gauge transformations
of $\tphi,\tA$ are nontrivial.
Away from $\tphi = 0$, the transformations
will change.  We will assume a more
general form good far from the open string vacuum:
\eqn\gengauge{
\eqalign{
    \delta \tphi = - f(\tphi) \tA_\mu \p_\mu \Lambda \ldots\cr
    \delta \tA_\mu = g(\tphi) \p_\mu \Lambda \ .
}}
Let us assume that the gauge transformations
acting on the BSFT variables $\phi, A$ are
of the standard form:
\eqn\standardgi{
\eqalign{
    & \delta \phi = 0\cr
    & \delta A = \p_\mu \Lambda\ ,
}}

Under these gauge transformations, the potential energy
$V(\tphi)$ is not gauge invariant.  Consider
small fluctuations of the gauge field.  To lowest order
in $\Lambda$ and $\tA$, the gauge-invariant completion is:
\eqn\invact{
    \CV(\tphi, \tA) = V(\tphi) + \half \frac{f}{g}
        \p_{\tphi} V \tA^2\ + \ldots
}
The field redefinitions required to
relate the gauge invariances \gengauge,\standardgi\
at this lowest order in $\tA$ are:
\eqn\redef{
\eqalign{
        &\phi = \tphi + \half
                \frac{f}{g}(\tphi) \tA^2\cr
        &A = \frac{\tA}{g(\phi)}
}}
At the minimum of $V(\tphi)$, the mass term vanishes
unless $g(\tphi)$ has a zero at the same order
as $\p_{\tphi} V$.  In this case, the field
redefinitions are singular precisely at the
closed string vacuum.  It is not clear if one
set of coordinates is more natural.

We can turn this story around and begin
with the variables $\phi, A$.
$V(\phi)$ has been calculated exactly
(at tree level) in BSFT.
If we expand $V(\phi)$ in $\tA^2$ we will find
the same mass term as above.  We will
also find that higher-order terms
in $\tA^2$ will diverge, because of the
pole in \redef\ and because
of the divergences in higher derivatives of
\canonpot.  The small fluctuation
expansion breaks down badly.  In order to
understand the relationship between
the various calculations, we need
a better handle on interaction terms
(and on higher-derivative terms) in the
cubic string field theory.

\listrefs
\end